\documentclass[aps,preprint,showkeys,nofootinbib]{revtex4-1}
\usepackage[utf8]{inputenc}
\usepackage{amsmath}
\usepackage{amsfonts}
\usepackage{amssymb}
\usepackage{hyperref}
\usepackage{color}
\usepackage{bbm}
\usepackage{epsfig}

\begin{document}
\author{Yago Ferreiros}
\email{yago.ferreiros@imdea.org}
\affiliation{IMDEA Nanociencia, Faraday 9, 28049 Madrid, Spain}
\author{Karl Landsteiner}
\email{karl.landsteiner@csic.es}
\affiliation{Instituto de F\'isica Te\'orica UAM/CSIC, c/Nicol\'as Cabrera 13-15, Universidad Aut\'onoma de Madrid, Cantoblanco, 28049 Madrid, Spain}

\title{On chiral responses to geometric torsion}
\date{\today}

\begin{abstract}
We show that geometric torsion does not lead to new chiral dissipationless transport effects. Instead apparent response to torsion can be viewed as a manifestation of the chiral vortical effect.
\end{abstract}
\preprint{IFT-UAM/CSIC-20-164}
\keywords{Weyl fermions, Torsion, Transport}

\maketitle

\section{Introduction}

Torsion is a geometric property of space-time describing the failure of a parallelogram spanned by two vectors to form a closed curve. In a perhaps more intuitive wording, vectors are twisted when parallel transported around a curve in a differential manifold with torsion \cite{HHK76}. While there is seeemingly no experimental evidence for torsion in the space-time of our universe, there are extensions of general relativity that include it, such as the Einstein-Cartan theory \cite{C22}\footnote{Recently Einstein-Cartan theory has also been considered for models of Higgs inflation in cosmology \cite{Shaposhnikov:2020gts}}. In materials, however, torsion does exist in the form of e.g. dislocations, and has been discussed in the context of Weyl semimetals \cite{Parrikar:2014usa,LS14,Z15,YCH16,Y16,SF16,CZ17}, topological insulators \cite{Parrikar:2014usa,HLF11,Hughes:2012vg}, graphene \cite{JCV10}, or Helium-3 \cite{IMT19}. Recently, there has been an increasing interest in condensed matter in what has been called (perhaps misleadingly) torsional (or Nieh-Yan) anomaly \cite{O182,NY282,O283,Y88,Y96,CZ97,Parrikar:2014usa,Hughes:2012vg,Ferreiros:2018udw,HLZ19,Huang:2019haq,Nissinen:2019wmh,Nissinen:2019mkw,N20,HH20,LN20,HHS20}. It represents the non-conservation of the chiral current of Weyl fermions given by the so called Nieh-Yan tensor \cite{Nieh:2007zz}. Due to dimensional reasons, the Nieh-Yan tensor has to be accompanied by a dimensionfull quantity in the anomaly equation. In vacuum this can only be the cutoff making it very ambiguous from the point of view of relativistic quantum field theory. In the condensed matter setting, however, there is a natural cutoff that could provide such a scale so that the Nieh-Yan term can arise \cite{Parrikar:2014usa,Hughes:2012vg,Ferreiros:2018udw,HLZ19,N20,HH20,LN20,HHS20}. It has also been argued that temperature could do the trick, claiming that there is a universal Nieh-Yan contribution to the anomaly which is proportional to temperature \cite{Huang:2019haq,Nissinen:2019wmh,Nissinen:2019mkw,LO20}. This is in line with some of the recent works which have studied chiral torsional transport at finite temperature \cite{KZ18,IY19,IQ20}.

Much of the motivation in looking for chiral torsional transport comes from the well established chiral magnetic  and chiral vortical effects (see \cite{landsteiner2013anomalous,Kharzeev:2013ffa,Kharzeev:2015znc,Landsteiner:2016led} for reviews). Both of them are direct consequences of anomalies \cite{FKW08, Son:2009tf, Alekseev:1998ds, Giovannini:1997eg, Landsteiner:2011cp, Jensen:2012kj, Golkar:2015oxw, Stone:2018zel}. 
In this letter we derive Kubo-formulas for chiral torsional transport, and show that in fact there is no dissipationless chiral electric transport as a response to torsion (to linear order in the torsion tensor), and that previously derived and discussed results in the literature can be understood as just a manifestation of the chiral vortical effect (CVE)\footnote{In contrast  to geometric torsion,  twisting a crystal deforms the crystal lattice and can lead to manifestations of the CME \cite{Gao:2020gcf}}. A similar conclusion has been reached in a recent model based on the gauge-gravity duality \cite{Gallegos:2020otk}. Chiral torsional effects in fluids with anomalies have also been recently investigated in \cite{Manes:2020zdd}. The treatment there relies on writing the contorsion tensor in terms of a dual axial vector field and does not directly address the question of transport induced by a genuine torsional anomaly given by the Nieh-Yan tensor.

\section{Torsion and possible dissipationless transport terms}

On a formal level, the torsion tensor is given by the anti-symmetric part of the affine connection
\begin{equation}
\theta^\lambda_{\mu\nu} = \Gamma^\lambda_{\mu\nu} - \Gamma^\lambda_{\nu\mu}\,.
\end{equation} 
In the following we will often employ differential form notation. We therefore introduce the vielbein $e^a = e^a_\mu dx^\mu$ and the spin connection $\omega^a\,_b = \omega_\mu\,^a\,_b dx^\mu$. 
The vielbein has a unique dual and obeys the relations
\begin{equation}
e^a_\mu E^\mu_b = \delta^a_b~~~,~~~e^a_\mu e^b_\nu \eta_{ab} = g_{\mu\nu}\,,
\end{equation}
where the tangent space metric is $\eta = \mathrm{diag}(+,-,-,-)$. The epsilon tensor is $\epsilon_{\mu\nu\rho\lambda} = e^a_\mu e^b_\nu e^c_\rho e^d_\lambda \epsilon_{abcd}$ where $\epsilon_{abcd}$ is taken as the totally antisymmetric symbol with 
$\epsilon_{0123}=1$.
Torsion and curvature are 
then the two forms
\begin{align}\label{eq:deftorsion}
\theta^a &= de^a + \omega^a\,_b \wedge e^b\,,\\
\label{eq:defcurv}
R^a\,_b &= d\omega^a\,_b + \omega^a\,_c \wedge \omega^c\,_b \,.
\end{align}
Assuming the vielbein to be covariantly constant gives a unique relation between all the components of the vielbein, torsion and spin connection $\omega^{abc} = E^{\mu a} \omega_\mu^{bc}$
\begin{equation}
\omega^{abc} = \frac{1}{2} E^{\mu a} E^{\nu b} (\partial_\mu e^c_\nu - \partial_\nu e^c_\mu - \theta_{\mu\nu}^c) - (acb) - (bca)\,,
\end{equation}

We will investigate if the presence of geometric torsion can lead to chiral transport analogous to the well known chiral magnetic and chiral vortical effects, CME and CVE respectively. 
We will consider minimally coupled spinors with action
\begin{equation}
S = \frac{i}{2}\int   \left( \bar\psi \star\gamma \wedge  \nabla\psi + \nabla \bar\psi \star\gamma \psi\right)\,.
\end{equation}
where
\begin{equation}
\nabla \psi = ( d  - i A - i A^5\gamma_5 -\frac{i}{2} \omega\,^{ab} \Sigma_{ab})\psi\,.
\end{equation}
and $\gamma = \gamma_a e^a$ and $\star$ denotes the Hodge star. On a $p$-form $\alpha$ the Hodge star acts as $\star\alpha = \frac{1}{(D-p)!p!} 
\epsilon_{\mu_1\dots\mu_{D-p}\nu_1\dots\nu_p}  dx^\mu_1\wedge\dots\wedge dx^\mu_{D-p}\alpha^{\nu_1\dots\nu_p}$.
The matrices $\gamma^a$ are the Dirac matrices in a convenient representation, $\Sigma_{ab} = \frac{i}{4} [\gamma_a,\gamma_b] $ are the generators of the Lorentz algebra in spin representation, and  $\gamma_5 = -i \star ( \gamma \wedge \gamma \wedge\gamma \wedge \gamma)$.
Projection onto right- and left-handed chiral fermions can be done with the chiral projectors $P_{R,L} = \frac 1 2 (1\pm \gamma_5)$. 
We define the current and axial current as three forms by
\begin{align}\label{eq:defJ}
\delta S &= \int \delta A\, J \,,\\
\label{eq:defJ5}
\delta S &= \int \delta A^5\, J_5\,.
\end{align}
These are the Hodge duals to the currents $J_{(5)}^\mu$. The spin current is similarly defined by
\begin{equation}\label{eq:ensemble}
\delta S = \int  \frac{1}{2}\delta\omega^{ab} S_{ab}\,.
\end{equation}
In our free fermion model these currents are
\begin{align}
J &= \bar \psi \star\gamma \psi\,,\\
J_5 &= \bar \psi \star \gamma \gamma_5 \psi \,,\\
S_{ab} &= \frac{1}{2} \bar\psi \{ \star\gamma, \Sigma_{ab} \} \psi\,.
\end{align}
Because of the identity 
\begin{equation}\label{eq:id}
\{\gamma_a, [\gamma_b,\gamma_c]\} = 4 i \epsilon_{abcd} \gamma_5 \gamma^d
\end{equation}
the components of the spin current are given by the components of the axial current. We note however that the conservation equations of axial and spin
current are independent of each other. On the quantum level the conservation equation might be broken of course by anomalies.
It is well known that CME and CVE are consequences of chiral anomalies.

For torsion a candidate anomaly term is the so-called Nieh-Yan tensor
\begin{equation}\label{eq:NiehYan}
d(e^a\wedge \theta_a) = \theta^a \wedge \theta_a - R_{ab}\wedge e^a\wedge e^b\,.
\end{equation}
One immediate difference to the usual $F\wedge F$ or $\mathrm{tr} ( R\wedge R)$ terms appearing in the axial anomaly is that the Nieh-Yan term has only dimension two. This means that as a possible contribution to the anomaly it has to be multiplied with a dimensionful quantity $\Lambda$ in a candidate anomaly equation. In vacuum the only candidate for such a scale is the cutoff 
and the axial anomaly upon $\delta A^5 = d\lambda_5$ would then be
\begin{equation}
\delta S = -\int \lambda_5 d J_5 =  \int  \lambda_5[c_A F\wedge F + c_g R^{ab}\wedge R_{ab} + c_T \Lambda^2  (\theta^a\wedge\theta_a - R_{ab}\wedge e^a \wedge e^b) ]\,.  
\end{equation}
We note that both fields, the torsion and the vielbein, are well defined tangent space tensor valued forms. 
The Nieh-Yan tensor is therefore a total derivative of well defined tensor fields. This is not true for the second Chern class $F\wedge F$ or the Pontryagin class $R^{ab}\wedge R_{ab}$. The latter two can not be written as derivatives of well defined tensors but the corrsponding Chern-Simons terms show explicit dependence on the local value of the connections $A$ or $\omega^{ab}$.
It seems therefore that the Nieh-Yan term could be removed by a suitable counterterm if $e^a\wedge \theta_a$ is simply interpreted as a contribution to the current. This still leaves the problem that the coefficient of such a contribution to the current depends on the cutoff scale $\Lambda$ and seems therefore highly ambigous from the standpoint of relativistic quantum field theory. That said we note that in a condensed matter context in which there is a natural cutoff it has been argued that indeed such terms do arise in \cite{Hughes:2012vg,Parrikar:2014usa}.

For this reason the notion of torsional anomaly is a rather complicated one. Instead of basing our 
discussion on the conservation equations we will directly calculate possible torsional analogues of CME and CVE via Kubo formulas. We will  investigate if (dissipationless) currents proportional to the torsion can be generated if physically well defined low energy scales are present in the theory. This makes it natural to consider the theory at finite temperature and chemical potential. Since we are after dissipationless currents analogous to the chiral magnetic and chiral vortical effects we also impose the restriction of time reversal invariance. This has been shown to be a very useful approach to CME and CVE in \cite{Kharzeev:2011ds}.
In order to do this we need a suitable notion of time reversal. An equilibrium state of a quantum statistical system is defined by the temperature and the chemical potentials for the conserved charges, in which we include the momenta. Therefore the state is characterized by $T$, $\mu$, and $u^a$. The latter encode the chemical potentials for the momenta. We can think of it as a four velocity normalized to $u_a u^a = 1$. Thus only three components are independent corresponding to the three momenta. 
The general statistical operator is
\begin{equation}
\rho = \exp[-\frac 1 T (u^a P_a - \mu_f Q_f) ]\,.
\end{equation}
The index $f$ runs over the different conserved $U(1)$ charges, e.g vector and axial charge in the case of a Dirac fermion and $P^a$ is the four momentum
\begin{align}
Q_f &  = \int_\Sigma J_f \,\\
P_a &  = \int_\Sigma T_a \,,
\end{align}
Here $J_f$ are the conserved current three forms, $T_a$ is the energy-momentum three form (see (\ref{eq:defemtensor})) and $\Sigma$ is an everywhere spacelike hypersurface. 
We define time reversal by  $\mathcal{T}: u_a \rightarrow -u_a$. We can assign consistent time reversal eigenvalues to the fields as follows

\begin{table}[h]
\begin{tabular}{c|c|c|c|c|c|c|c|c|c|c|c|}
   & $A_{(5)}$ & $e^a$ & $\omega^{ab}$ & $d$  & $J_{(5)}$  & $S_{ab}$ & $u_a$ & $\mu$ & $T$  & $\star$\\ \hline
$\mathcal{T}$: & $-1$      & $+1$  & $+1$          & $+1$ & $+1$       & $-1$     & $-1$  & $+1$  & $+1$ & $-1$
\end{tabular}\caption{Action of time reversal}\label{tab:Taction}
\end{table}
The action on the spinors depends on the representation of the Dirac matrices. Since time reversal acts anti-unitary taking the imaginary unit $i\rightarrow -i$, it is easiest to chose momentarily a Majorana representation in which all $\gamma_a$ matrices are purely imaginary. Taking then $\mathcal{T}: (\bar\psi,\psi) \rightarrow (\bar\psi,\psi)$ leaves the action invariant\footnote{This is not the usual version of time reversal discussed in textbooks. In order to connect to the usual one we note that our $\mathcal{T}$ in flat spacetime takes $e^0_t = 
\delta^0_t \rightarrow -\delta^0_t$, i.e. it changes the background. To compensate for this one can combine the $\mathcal{T}$ operation just defined with an $O(1,3)$ transformation $\Lambda^a\,_b = \mathrm{diag}(-1,1,1,1)$ acting on the tangent space indexes. This combined operation $\mathcal{T}'$ is effectively the usual time reversal operation in Minkowski spacetime. Its matrix representation $T'$ acts as $\mathcal{T}'(\gamma^0) : T'\gamma^{0*} T'^{-1} = \gamma^0$ and $\mathcal{T}'(\gamma^{i}): T' \gamma^{i*} T'^{-1} = -\gamma^i$. In Majorana representation $T'=\gamma^0\gamma_5$ as discussed e.g. in \cite{Peccei:1998jv}.}.
This allows to classify possible response terms in the current $J$ according to their eigenvalues under the $\mathcal{T}$ operation.
As an example we take the possible responses due to presence of a gauge field $A$. Since the current should be a gauge invariant operator the response has to be expressed in terms of the field strength $F=dA$. To linear order there are two possible terms
\begin{equation}
J = \sigma_\Omega u\wedge \star F + \sigma_\mathrm{cme} u\wedge F\,.
\end{equation}
The first term is $\mathcal{T}$ odd and represents the usual Ohmic conductivity whereas the second term is $\mathcal{T}$ even and represents the CME. 
In the same way we can now classify possible $\mathcal{T}$ even response terms that depend at most linearly on torsion or vorticity $du$. We find
\begin{equation}\label{eq:constJ}
J = \rho \star u + c_V u\wedge du + c_T^\parallel u_a u_b \theta^a e^b + c^\perp_T P_{ab} \theta^a e^b,
\end{equation}
where we defined the projector $P_{ab} = \eta_{ab}-u_a u_b$.
A formally completely identical expression can be written down for the axial current. As we will see it is necessary also to consider the convective term $\rho\star u$ of zero order in derivatives. The reason is that torsion contains both zero and first order terms in derivatives, and this is important to take into account in the following.
The chiral vortical coefficient $c_V$ can be calculated via Kubo formulas by assuming a pseudo-Riemannian spacetime with vanishing torsion \cite{Amado:2011zx}. In the normalization of eq. (\ref{eq:constJ}) the chiral vortical coefficients for Weyl fermions are   \cite{Landsteiner:2011cp}
\begin{equation}
c_V = \pm \left( \frac{\mu_{R,L}^2}{8\pi^2} + \frac{T^2}{24}\right)\,.
\end{equation}  
The corresponding expressions for vector and axial current are obtained by taking the sum and the difference.

\section{Derivation of Kubo formulas}

There are now in principle two ways to derive Kubo type relations for the unknown transport coefficients $c^\parallel_T$ and $c^\perp_T$. 
The first possibility is to take a flat background geometry $e^a=dx^a$ and $\omega^a\,_b=0$, and then introduce a suitable and infinitesimally small component of the spin connection $\delta \omega_\mu\,^a\,_b$ while keeping the vielbein fixed. 

Since the spin current is determined by the components of the axial current, we also note that switching on the spin connection is equivalent to switching on certain components of an axial background field $A^5$. In particular we consider the response to the spin connection components $\delta \omega_x\,^t\,_y=\delta\omega_x\,^y\,_t$, $\delta \omega_t\,^x\,_y=- \delta \omega_t\,^y\,_x$,
and $\delta\omega_x\,^y\,_z = -\delta\omega_x\,^z\,_y$. Because of the identity (\ref{eq:id}) this has the same effect as switching on the axial gauge field perturbations $\delta A_z^5$, $\delta A^5_t$. The response can therefore be expressed as
\begin{align}
J^z &= \frac{1}{2}(c^\parallel_T + c^\perp_T) \delta A^5_z\,,\\
J^z &= - c^\perp_T \delta A^5_z\,,\\ \label{eq:kubosus}
J^t &= (\frac{\partial \rho}{\partial \mu_5} -  c^\perp_T) \delta A^5_t\,.
\end{align}
In the last equation we have taken into account that a variation of the temporal component also shifts the chemical potential $\delta \mu = \delta A_t$. Therefore the zero order term in (\ref{eq:constJ}) also contributes to the response (\ref{eq:kubosus}).
As usual we interpret these expressions as the expectation values of the current an charge.
Differentiating with respect to the axial gauge field components therefore gives two point correlation functions
\begin{align}
\langle J^z J^z_5 \rangle &= \frac{1}{2}(c^\parallel_T + c^\perp_T)=-c_T^\perp\,,\\
\langle J^t J^t_5 \rangle &= \frac{\partial \rho}{\partial \mu_5}-c_T^\perp\,.
\end{align}
Instead of the vector current $J$ we could also have considered the axial current $J_5$. Note that the charge susceptibility, $\partial \rho/\partial \mu_{5}$, is defined precisely by $\langle J^t J_{5}^t\rangle$ evaluated in the thermodynamic limit, which is to say that one first takes the frequency to zero and then the momentum. The correlator is quite standard to evaluate (see for example \cite{le2000thermal})
\begin{equation}
    \langle J^t J_{5}^t\rangle=\frac{2\mu\mu_5}{\pi^2}(1-I(\omega,\vec{p})),
\end{equation}
\begin{equation}
    \langle J^t_5 J_{5}^t\rangle=(\frac{\mu^2+\mu_5^2}{\pi^2}+\frac{T^2}{3})(1-I(\omega,\vec{p})),
\end{equation}
where
\begin{equation}
    I(\omega,\vec{p})=\int\frac{d\Omega}{4\pi}\frac{i\omega}{i\omega+\vec{p}\cdot\hat{q}},
\end{equation}
with $\hat{q}=\vec{q}/q$. In the thermodynamic limit $I=0$ and one gets $\langle J^t J_{5}^t\rangle=\partial\rho/\partial\mu_5=2\mu\mu_5/\pi^2$ and $\langle J^t_5 J_{5}^t\rangle=\partial\rho/\partial\mu=(\mu^2+\mu_5^2)/\pi^2+T^2/3$. Because we are looking for time-reversal invariant response terms similar to CME and CVE, this is precisely the limit we have to impose. It therefore follows that
\begin{align}\label{eq:zero}
c^\perp_T=c^\parallel_T =0\,.
\end{align}
To cross-check this result, we can also evaluate $\langle J^z_{(5)} J^z_5 \rangle$ (again from \cite{le2000thermal})
\begin{equation}
    \langle J^z J_{5}^z\rangle=\frac{2\mu\mu_5}{\pi^2}I(\omega,\vec{p}),
\end{equation}
\begin{equation}
    \langle J^z_5 J_{5}^z\rangle=(\frac{\mu^2+\mu_5^2}{\pi^2}+\frac{T^2}{3})I(\omega,\vec{p}),
\end{equation}
which indeed vanishes in the thermodynamic limit and confirms result \eqref{eq:zero}.

Alternatively we could also derive this result by considering a spacetime background with vanishing spin connection and a vielbein of the form $e^a = dx^a + h^a_\mu dx^\mu$, and work to linear order in $h^a_\mu$. 
The energy and momentum currents are defined as
\begin{equation}\label{eq:defemtensor}
\delta S = \int T_a \delta e^a\,.
\end{equation}
Again this is the  Hodge dual to $T^\mu_a$. 
For the minimally coupled theory 
\begin{equation}
T_a = \frac{i}{4}\epsilon_{abcd} e^b \wedge e^c \wedge ( \bar\psi \gamma^d \nabla \psi - \nabla\bar\psi \gamma^d \psi)\,.
\end{equation}
We note that this definition distinguishes between the energy current $T^\mu_t$ and the energy and momentum densities $T^t_a$. Switching on a perturbation $\delta e^t = h^t_x(y) dx$ leads therefore to an insertion of the  $x$-component of the energy current, whereas the perturbation $\delta e^x = h^x_t(y) dt$ couples to the momentum density in $x$-direction.
The perturbation $\delta e^t = h^t_x dx$ modifies the one form $u=u_a e^a$ and therefore enters in the chiral vortical effect. 
Momentum is a conserved charge and an ensemble in equilibrium is also defined by the corresponding chemical potential, the fluid velocity $u^a$. Therefore switching on the perturbation $h^x_t dt$ leads also to a shift in the fluid velocity as is clear from
(\ref{eq:ensemble}) and (\ref{eq:defemtensor}). This is completely analogous to the increase in the chemical potential due to a variation in the temporal component of a gauge field $\delta \mu = \delta A_t$, and thus $\delta u^x = \delta e^x_t$. The corresponding response to both types of perturbations is therefore a linear combination of chiral vortical effect and a possible torsion response
\begin{align}
J &= -(c_v + c^\parallel_T) dh^t_x dt dx\,,\\
J &= (c_v + c^\perp_T) dh^x_t dt dx \,.
\end{align} 
We get correlation functions of current and energy current and momentum density by taking $h^t_x$ and $h^x_t$ to depend only on $y$. Then
 \begin{align}
\langle J^z T^{xt}\rangle &= (c_v + c^\parallel_T) i p_y\,,\\
\langle J^z T^{tx}\rangle &= (c_v + c^\perp_T) i p_y\,.
\end{align}
Now we note that these exact correlators have been calculated in \cite{Landsteiner:2011cp} with the result that both exactly equal the chiral vortical coefficient $c_v$. Therefore we find again the result (\ref{eq:zero}).

\section{Conclusions}

We have derived all possible 
time reversal invariant torsional responses (to linear order) in the electric and axial currents of a Dirac fermion minimally coupled to geometric torsion. We have shown that all such coefficients vanish. This implies that there is no universal chiral transport under torsional fields. This clarifies some discussions in previous literature, where the responses found can be explained as responses to vorticity through the CVE in a generalized background geometry. Indeed there is an ambiguity in formulating the CVE response. By defining the tangent space valued vorticity one form $\Omega_a = Du_a$ 
and noting that $du = du_a\wedge e^a + u_a de^a + u_a \omega^a\,_b\wedge e^b - u_a \omega^a\,_b\wedge e^b = (Du_a)\wedge e^a + \theta^a u_a$ the CVE can also be written as 
\begin{equation}
    J = c_v u\wedge du= c_v u \wedge\left( \Omega_a \wedge e^a  +  \theta^a u_a\right)\,.
\end{equation}
Formally the absence or presence of response to torsion depends therefore on the definition of vorticity. 
The transport coefficient is however always given by the chiral vortical one.  

Finally we would like to comment on the generality of our results. We considered a minimally coupled spinor in which the identity (\ref{eq:id}) played a major role in identifying the spin current with the axial current. This means that the spin connection is always projected onto its totally anti-symmetric part and enters the theory only through the combination $\frac 1 2 \epsilon^{abcd} \omega_{abc}$ in which it couples to the axial current. Our result on the absence of the torsion coefficients $c_T^{(\parallel,\perp)}$ has been obtained in two independent ways. The derivation via the vielbein couplings does not depend on (\ref{eq:id}) and is therefore generally valid.

\newpage

\acknowledgments
K.L. has been supported by  Agencia Estatal de Investigaci{\'o}n IFT Centro de Excelencia Severo Ochoa
SEV-2016-0597, and by the grant PGC2018-095976-B-C21 from MCIU/AEI/FEDER, UE. \\
Y.F. acknowledges financial support through the Programa de Atracción de Talento de la Comunidad de Madrid, Grant No. 2018-T2/IND-11088. \\
We thank M. Chernodub, M. Garbiso, A. Grushin, U. Gurosy, M. Kaminski, J. Ma{\~n}es, J. Nissinen, F.Pena-Benitez, M. Valle, M.A. Vazquez-Mozo, G. Volovik and M.A.H. Vozmediano for discussions and helpful comments on the manuscript.

\appendix

\bibliography{TorsionTransport}{}

\end{document}